\documentclass[twocolumn,3p,final,times]{elsarticle}

\usepackage{graphicx,amsmath,amssymb}
\usepackage{hyperref}
\usepackage{breakurl}
\usepackage{color}
\usepackage{verbatim}
\usepackage{multirow}
\usepackage{tabularx}
\usepackage{slashed}
\usepackage{soul}
\usepackage{ulem}

\newcommand{\ie}{\textit{i.e.}}

\newcommand{\be}{\begin {equation}}
  \newcommand{\ee}{\end{equation}}
\newcommand{\bi}{\begin{itemize}}
  \newcommand{\ei}{\end{itemize}}
\newcommand{\bea}{\begin {eqnarray}}
  \newcommand{\eea}{\end{eqnarray}}
\newcommand{\braket}[2]{\bra{#1}\,#2\rangle} % Dirac inner product
\newcommand{\bra}[1]{\langle\,#1\,|}          % Dirac bra
\newcommand{\ket}[1]{|\,#1\,\rangle}          % Dirac Ket
\newcommand{\ud}{\mathrm{d}}

\newcommand{\LCm}{{\scriptscriptstyle -}} %LC supersripts 
\newcommand{\LCp}{{\scriptscriptstyle +}}

\begin{document}

\begin{frontmatter}

  \title{Non-perturbative quantum time evolution on the light-front}

  \author[XZ]{Xingbo Zhao}
  \ead{xbzhao@iastate.edu}

  \author[AI]{Anton Ilderton}
  \ead{anton.ilderton@chalmers.se}

  \author[PM]{Pieter Maris}
  \ead{pmaris@iastate.edu}

  \author[JV]{James P. Vary} 
  \ead{jvary@iastate.edu}

  \address{$^1$Department of Physics and Astronomy, Iowa State University, Ames, Iowa 50011, USA}
  \address[AI]{Department of Applied Physics, Chalmers University of Technology, SE-412 96 Gothenburg, Sweden}

\begin{abstract}
This paper introduces ``time-dependent basis light-front quantization", which is a covariant, nonperturbative, and first principles numerical approach to time-dependent problems in quantum field theory. We demonstrate this approach by evaluating photon emission from an electron in a strong, time and space dependent external field. We study the acceleration of the electron in combination with the evolution of the invariant mass of the electron-photon system, in real time and at the amplitude level. Successive excitations of the electron-photon system are evident.

  \end{abstract}

\end{frontmatter}

% \pacs{11.10.Ef, 11.15.Tk, 12.20.Ds}
% 
\paragraph{Introduction} Recent interest in strong-field dynamics covers topics such as (a) the observed anomalous enhancement of lepton production in ultrarelativistic nuclear collisions at RHIC~\cite{Adare:2009qk}, (b) a prediction for photon yield depletion at the LHC~\cite{Tuchin:2010gx},   (c) effects of strong magnetic fields on quantum chromodynamics (QCD)~\cite{Chernodub:2011mc,Bali:2011qj,Basar:2011by,Tuchin:2012mf}, and (d) studying both quantum electrodynamics (QED) and beyond-standard-model physics using next-generation lasers~\cite{DiPiazza:2011tq, Heinzl:2011ur, Jaeckel:2010ni}. This broad range of frontier applications points to the importance of developing new methods for solving time-dependent problems in quantum field theory (QFT), in the nonperturbative regime.

Nonperturbative QFT calculations present a significant challenge even for stationary state solutions.  The need to retain covariance and to develop numerical methods of sufficiently high precision, at practical computational cost, poses immense challenges.  A well-suited, flexible, and first-principles framework for this problem is provided by the Hamiltonian light-front formalism~\cite{Brodsky:1997de}, in which the theory is quantized on the light-front. The formalism is Lorentz frame independent and all observables are, in principle, accessible from the full time-dependent wavefunction.

Within the Hamiltonian light-front formalism, ``Basis Light-Front Quantization'' (BLFQ)~\cite{Vary:2009gt}, provides a nonperturbative calculational method for stationary states of QFT~\cite{Vary:2009gt,Honkanen:2010rc}. BLFQ diagonalizes the full QFT Hamiltonian and yields the physical mass eigenstates and their eigenvectors. This approach offers opportunities to address many outstanding puzzles in nuclear and particle physics~\cite{Brodsky:2011vc,Brodsky:2012rw}.

Here we introduce the logical extension of BLFQ, called {\it time-dependent} Basis Light-Front Quantization (tBLFQ), which provides the fully quantum, real time\footnote{Since we use light-front quantization, ``time'' will be synonymous with ``light-front time'' throughout this paper.} evolution of a chosen initial state under the influence of both quantum effects and applied background fields with arbitrary spacetime-dependence. We focus here on the principles of the method, for all details see~\cite{Zhao:2013cma}.

We apply tBLFQ to ``strong field QED'', in which the background models the high-intensity fields of modern laser systems, which routinely reach intensities of $10^{22}$ W/cm$^2$ and offer prospects for investigating effects such as vacuum birefringence~\cite{Heinzl:2006xc}. We will study one of the simplest background-field stimulated processes, namely photon emission from an electron which is excited by a background field.  (In a plane-wave background, this well-studied process goes by the name ``nonlinear Compton scattering''~\cite{Nikishov:1963,DiPiazza:2011tq}) A major challenge in strong-field QED (see the review~\cite{DiPiazza:2011tq}) is obtaining analytic results beyond the approximation of constant backgrounds or backgrounds with one-dimensional field inhomogeneities, in the regime that the background is strong. This poses however no issue in our approach (see also \cite{Hebenstreit:2011wk,DiPiazza:2012ur}), and to demonstrate this we will consider backgrounds with two-dimensional inhomogeneities, depending on both space and time.

\paragraph{Light-front dynamics} Physical processes in light-front dynamics are described in terms of light-front time $x^+\equiv x^0+x^3$ as well as transverse co-ordinates $x^\perp$=\{$x^1$, $x^2$\} and the longitudinal direction $x^-$=$x^0-x^3$ with $-L<x^\LCm<L$. We therefore begin with an electron which, at light-front time $x^+{=}0$, enters the laser field.  The electron can be both accelerated (invariant mass unchanged) and excited (invariant mass changed) by the background, with excitation producing electron--photon final states.

\begin{figure}[!t]
  \centering
  \centering\includegraphics[width=0.45\textwidth]{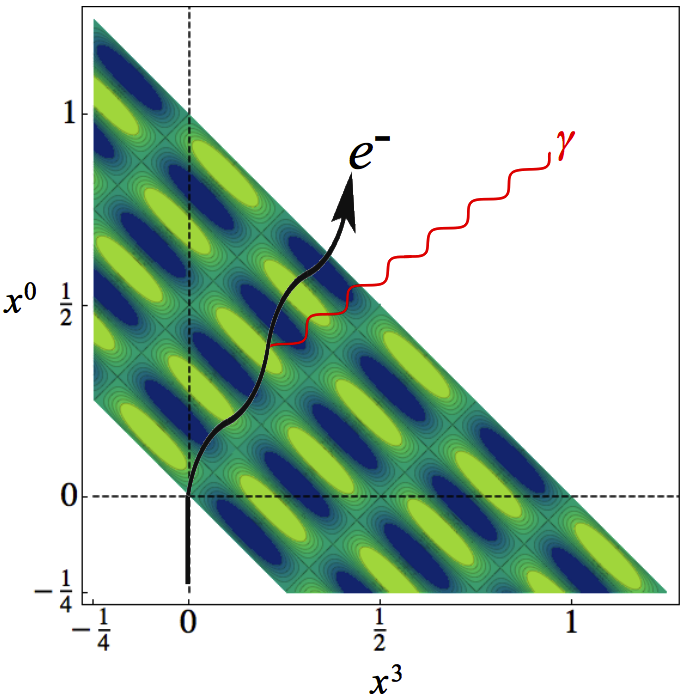}
  \caption{\label{fig:nCs} Schematic density plot of the two-dimensional field $\mathcal{F}_{+-}$, describing a finite pulse traveling along the $z$-direction. An electron entering the field is accelerated, and emits a photon.}
\end{figure}

While tBLFQ can accommodate arbitrarily complex background fields, as a first step we consider a simple two-dimensional field. Our background is turned on for a finite light-front time $\Delta x^+$, during which it depends on both $x^+$ and $x^-$ as follows:
\be
\mathcal{F}_{+-} = \frac{1}{2} E_0\sin\left(l_-x^-\right) \sin\left(l_+ x^+\right) \;.
\ee
This is an electric field pointing in the {3}-direction. It has periodic structure in the longitudinal and time directions  with frequencies $l_-$ and $l_+$ respectively. In the lab frame it describes a finite duration beam propagating along the $x^3$ direction, see Fig.~\ref{fig:nCs}. This background accelerates charges in the $x^-$ ($x^3$) direction as time $x^+$ ($x^0$) evolves. A convenient gauge potential is
\be
	e\mathcal{A}_+= a_0m_e\cos\left(l_-x^-\right)\sin\left(l_+ x^\LCp\right) \;,
\ee
where $e$ ($m_e)$ is the electron charge (mass) and we define the dimensionless parameter $a_0= e E_0/(2m_e l_\LCm)$. We are interested here in a ``proof of concept" demonstration of tBLFQ, not in phenomenology, so it is not an issue that our background does not obey Maxwell's equations in vacuum. In fact there are very few realistic, finite energy, closed-form solutions to Maxwell's equations available, see for example~\cite{Ivan}, but such fields are beyond the scope of this first investigation. Note also that whether the background obeys Maxwell or not has no impact on our methods. 

\paragraph{The tBLFQ approach} The principle behind our approach is simple: given an initial state, we will solve the light-front Schr\"odinger equation in order to find the state at later times.  The light-front Hamiltonian, $P^\LCm$, will contain two parts; $P^-_\text{QED}$ which is the {\it full}, interacting, light-front Hamiltonian of QED and explicitly time-dependent interactions $V(x^\LCp)$ introduced by the external field, so
\begin{align}
  \label{H_interact}
  P^-(x^+)=P^-_\text{QED}+V(x^+) \;.
\end{align}
In our case, the interaction term introduced by the background field $\mathcal{A}$ is 
  \be
  V(x^+)=\int\!\ud x^\perp \ud x^- \ e\bar\Psi \gamma^{\LCp}\mathcal{A}_\LCp \Psi\;,
  \ee
  where $\Psi$ is the fermion field operator.

Now, imagine that we had the eigenstates $\ket{\beta}$ of $P^\LCm_\text{QED}$. We would then be interested in the transitions between such states introduced by the background field interactions contained in $V$. This, and the explicit time dependence of our theory makes it natural to use an interaction picture, but {\it not} the usual ``free + interacting'' split of the Hamiltonian. We take our ``free" Hamiltonian to be $P^\LCm_\text{QED}$, and hence solve the Schr\"odinger equation in the form 
\be\label{Schro-int}
i\frac{\partial}{\partial x^+}\ket{\psi;x^+}_I= \frac{1}{2}V_I(x^+)\ket{\psi;x^+}_I \;,
\ee
in which $V_I$, ``the interaction Hamiltonian in the interaction picture'', is (since $P^\LCm_\text{QED}$ is time-independent)
\be
V_I(x^+) = e^{\tfrac{i}{2}P^-_\text{QED}x^+}V(x^+)e^{-\tfrac{i}{2}P^-_\text{QED}x^+} \;.
\ee
The formal solution to (\ref{Schro-int}) is a time-ordered ($ \mathcal{T}_+$) series:
\begin{align}
  \label{i_evolve}
  \ket{\psi;x^+}_I  &= \mathcal{T}_\LCp e^{-\tfrac{i}{2}\int\limits_0^{x^+} V_I}\ket{\psi;0}_I \;,
\end{align}
in which we begin, at $x^+=0$, with a chosen initial state which will be a sum over QED eigenstates:
\begin{align}
  \label{initial_c}
  \ket{\psi;0}_I = \sum\limits_\beta \ket{\beta} c_\beta(0)\;.
\end{align}
We then expand the interaction picture state at later times as
\be\label{int-expand}
\ket{\psi;x^+}_I  := \sum_\beta c_\beta(x^+) \ket{\beta}.
\ee
Substituting this into (\ref{Schro-int}) yields a system of equations for the coefficients $c_\beta$, which we wish to solve. Clearly this is a nonperturbative approach, with the two challenges being to first construct the eigenstates of QED, and then to solve for their time evolution under the action of external fields. We make two corresponding approximations: ``basis truncation" and ``time-step discretization". These are the only approximations in our approach. Basis truncation means selecting a finite dimensional basis of states with which to represent the stationary state solutions in BLFQ. State vectors and time-evolution operators then become finite length column vectors and finite dimensional matrices respectively.  Time evolution is implemented numerically; we decompose the time-evolution operator in (\ref{i_evolve}) into small steps in light-front time $x^+$, introducing the step size $\delta x^+$,
\be
\label{sol_wave-eq_i_discrete}
\mathcal{T}_\LCp e^{-\frac{i}{2}\int\limits_0^{x^+} V_I}\rightarrow \big[1-\tfrac{i}{2}V_I(x^+_{n})\delta x^+\big] \cdots \big[1-\tfrac{i}{2}V_I(x^+_{1})\delta x^+\big] \;,
\ee
and we let these matrices act sequentially on the initial state. In this way we solve (\ref{Schro-int}) with initial conditions~(\ref{initial_c}). (To ensure numerical stability and preservation of the state norm, we implement (\ref{sol_wave-eq_i_discrete}) through the second order difference scheme MSD2~\cite{Askar:1978}.)

To implement BLFQ~\cite{Vary:2009gt,Honkanen:2010rc} for stationary states we must first consider the symmetries of QED. The
Hamiltonian $P^-_\text{QED}$ has, amongst others, the following three symmetries. 1) Translational symmetry in the $x^-$
direction, i.e.\ longitudinal momentum, $P^\LCp$, is conserved. 2) Rotational symmetry in the transverse plane i.e.\ longitudinal
projection of angular momentum, $J^3$ is conserved. 3) Lepton number conservation, i.e.\ charge, or net fermion number $Q$, is
conserved. The eigenspace of QED therefore breaks up into segments which are groups of eigenstates with definite eigenvalues $\{K, M_j, N_f\}$ for the operators \{$P^+, J^3, Q$\} respectively.

BLFQ uses a basis of states which has these three symmetries, i.e.\ each basis state $\ket\alpha$ obeys $\{P^+, J^3, Q\}
\ket{\alpha} = \{K, M_j, N_f\}\ket{\alpha}$, leading to a block-diagonal structure in $P^\LCm_\text{QED}$. The basis elements are
collections of Fock particle states, but these are not the usual momentum eigenstates of the free theory. In order to respect the
above symmetries, the Fock particles' transverse degrees of freedom are those of a two-dimensional harmonic oscillator (2D-HO), characterized by the mass scale\footnote{The transverse modes depend
    only on the combination $b = \sqrt{M \Omega}$, where $M$ and $\Omega$ are the usual HO mass and frequency
    parameters~\cite{Zhao:2013cma}.} $b$.   Each Fock particle carries four quantum numbers, $k,n,m,\lambda$. Here, $k$ labels the particle's longitudinal momentum $p^\LCp = 2\pi k/L$ where $k=1,2,3\ldots $ for bosons (neglecting the zero mode) and $k=\tfrac{1}{2},\tfrac{3}{2},\tfrac{5}{2}\ldots$ for fermions, signifying our convenient choice of boundary conditions. $n$ and $m$ label the quanta of the 2D-HO radial excitation and angular momentum, respectively. The Fock state carries the 2D-HO eigenenergy $E_{n,m}=(2n+|m|+1)\Omega$. Finally, $\lambda$ is the particle's helicity.

In BLFQ we expand the physical QED eigenstates, $\ket{\beta}$, as superpositions of the basis states, %
\begin{align}\label{phys-expand}
  \ket{\beta}=\sum_\alpha \ket{\alpha} \braket{\alpha}{\beta} \;,
\end{align}
in which the $\ket\alpha$ belong to the same segment as $\ket\beta$. We then diagonalize the Hamiltonian in this basis, which yields the coefficients $\braket{\alpha}{\beta}$. To do this calculation in practice we need to be able to diagonalize $P^\LCm_\text{QED}$. Since our aim is to implement tBLFQ numerically, we will clearly have to supply some truncation. To understand this truncation, we consider the form of QED eigenstates. The physical electron in QED is a composite field composed of the fermion and a cloud of photons~\cite{Dirac:1955uv}, with only the combination of these two being physical and observable~\cite{Lavelle:1995ty}. The physical electron therefore has components in all Fock-sectors with $N_f=1$, i.e.\
\begin{align}
  \label{state_expan_BLFQ_2}
  |e_\text{phys}\rangle=a|e\rangle+b|e\gamma\rangle+c|e\gamma\gamma\rangle+d|ee\bar{e}\rangle+\ldots\;,
\end{align}
where we see the bare fermion $\ket{e}$, the photon cloud $\ket{e\gamma}$, $\ket{e\gamma\gamma}$ and so on. In this paper, we make the simplest nontrivial truncation, which is to truncate our Fock-sectors to $\ket{e}$ and $\ket{e\gamma}$.

Truncation is also related to the unitarity of our approach. Note that, following (\ref{sol_wave-eq_i_discrete}), unitarity is conserved up to order $(\delta x^+)^2$ terms, but this is standard for numerical implementations. To this order, though, the Hermiticity of our Hamiltonian in the truncated subspace guarantees that total probability is conserved. In other words, any ``missing'' probability for reaching states outside our truncated basis are redistributed into probabilities for reaching retained states. In order for these probabilities to be accurate, transitions to states which are outside our basis should be small. It is thus necessary to consider the accessible kinematic regime, and which processes are/are not likely to occur, when choosing the basis truncation.

A complete specification of a BLFQ basis requires 1) the segment numbers $K,M_j,N_f$, 2) two truncation parameters, namely the
choice of which Fock sectors to retain, and the transverse truncation parameter $N_\text{max}$, the maximum total number of
oscillator quanta $2n+|m|$ for the Fock state and 3) the length $L$ of the longitudinal direction and parameter $b$ for the 2D-HO. (Renormalization in BLFQ, to be discussed in detail elsewhere, is achieved via a sector-dependent scheme~\cite{Karmanov:2008br,Zhao:2013xx,Chabysheva:2009ez}. This was successfully applied to a calculation of the electron's anomalous magnetic moment in~\cite{Zhao:2013xx};  the result agrees with the Schwinger value to within 1\%.)

While a single segment of states is sufficient to address, e.g.\ bound-state problems, the presence of the background field means that $P^-$ will not conserve longitudinal momentum ($K$) or longitudinal projection of total angular momentum ($M_j$). A background field can therefore cause transitions between different segments, and in time-dependent problems one must work with a basis which covers sufficiently many segments to capture the physics of interest. Our chosen background does not directly excite the transverse degrees of freedom, so we need only include segments with different total longitudinal momenta $K$.

We note that our choice of discrete basis preserves manifest rotational invariance in the transverse plane, but breaks
  manifest 3D rotational invariance. This is restored in the continuum limit, $N_\text{max}$ and $K$ to infinity, and with the
  Fock space truncation removed. It remains to be seen whether the truncated basis space can provide reasonably accurate solutions
  with rotational symmetry. Of course, to test 3D rotational symmetry requires dynamical operators in light-front
  dynamics. We also preserve boost invariance in the longitudinal direction, despite the Fock space truncation and
  limited mode expansions for the retained degrees of freedom. We may also, in principle, retain exact boost invariance in the
  transverse direction by factorizing the center-of-mass motion from the intrinsic motion~\cite{Zhao:2013xx,Maris:2013qma}.
Since our chosen background field does not excite the transverse degrees of freedom, we feel that transverse boost
    invariance is less important in the current application. Nevertheless, these issues are under investigation and will be
  reported elsewhere.

\paragraph{Photon emission}
For our process, the initial state is a single {\it physical} electron with longitudinal momentum $K_i$. This state can be
identified as the ``ground state'' of the QED Hamiltonian $P^-_\text{QED}$ in the segment $N_f{=}1$, $M_j{=}\tfrac{1}{2}$ and
$K$=$K_i$ (since there is no other state in that segment with a lower energy). In the tBLFQ basis, which is just the set of
eigenvectors of QED, this initial state is simply a column vector with a one for the lowest-energy eigenstate in the $K = K_i$ segment, and all other entries equal to zero.

Between each matrix multiplication in (\ref{sol_wave-eq_i_discrete}) we insert a (numerically truncated) resolution of the identity, so that the evaluation of (\ref{i_evolve}) amounts to the repeated computation of the overlaps $\bra{{\beta'}} V_{I} \ket{\beta}$. In order to calculate these overlaps in our numerical scheme, it is simpler to calculate them first in the BLFQ basis, and  then transform to the tBLFQ basis as follows:
\begin{align} \label{VLAS_trans}
  \bra{{\beta'}}V\ket{\beta}=\sum_{\alpha'\alpha} \braket{{\beta'}}{{\alpha'}} \bra{{\alpha'}}V\ket{{\alpha}} \braket{{\alpha}}{\beta} \;.
\end{align}
If $\bar\alpha$ and $\bar\alpha'$ label two Fock electron states, then one finds for example, with $k_\text{las} = l_\LCm L/\pi$,
\be
\bra{\bar{\alpha}'}V\ket{\bar\alpha} = m_e a_0 \sin\left(l_+ x^+\right)\big(\delta_{\lambda}^{\lambda'} \delta_{n}^{n'} \delta_{m}^{m'}\big)\ (\delta_{k}^{k'+k_\text{las}}+\delta_{k}^{k'-k_\text{las}}) \;,
\ee
in which $\delta^*_*$ is the Kronecker delta. The magnitude of the matrix element is proportional to the field intensity~$a_0$. 

The basis chosen here consists of three segments with
$K{=}\{K_i, K_i{+}k_\text{las}, K_i{+}2k_\text{las}\}$. In each segment we retain {\it both} the single electron (ground) and
electron-photon (excited) state(s). The initial state for our process is a single (ground state) electron in the $K{=}K_i$
segment. This basis allows for the ground state to be excited twice by the background (from the segment with $K$=$K_i$ through to
segment with $K_i$+2$k_\text{las}$). In this calculation, we take $K_i{=}1.5$ and $N_\text{max}{=}8$, with $a_0$=0.1,
$k_\text{las}{=}2$, $L=2\pi$~MeV$^{-1}$, $l_-{=}\frac{\pi}{L} k_{las}{=}1$~MeV (\ie, a period of $2\pi$~MeV$^{-1}$ in $x^-$ direction) and
$l_+{=}\frac{2\pi}{50}$~MeV (\ie, a period of 50~MeV$^{-1}$ in $x^+$ direction). We switch on the background at $x^+{=}0$ and
evolve the system with the step size of $\delta x^+{=}0.02$MeV$^{-1}$.
\begin{figure}[!t]
\centering\includegraphics[width=0.45\textwidth]{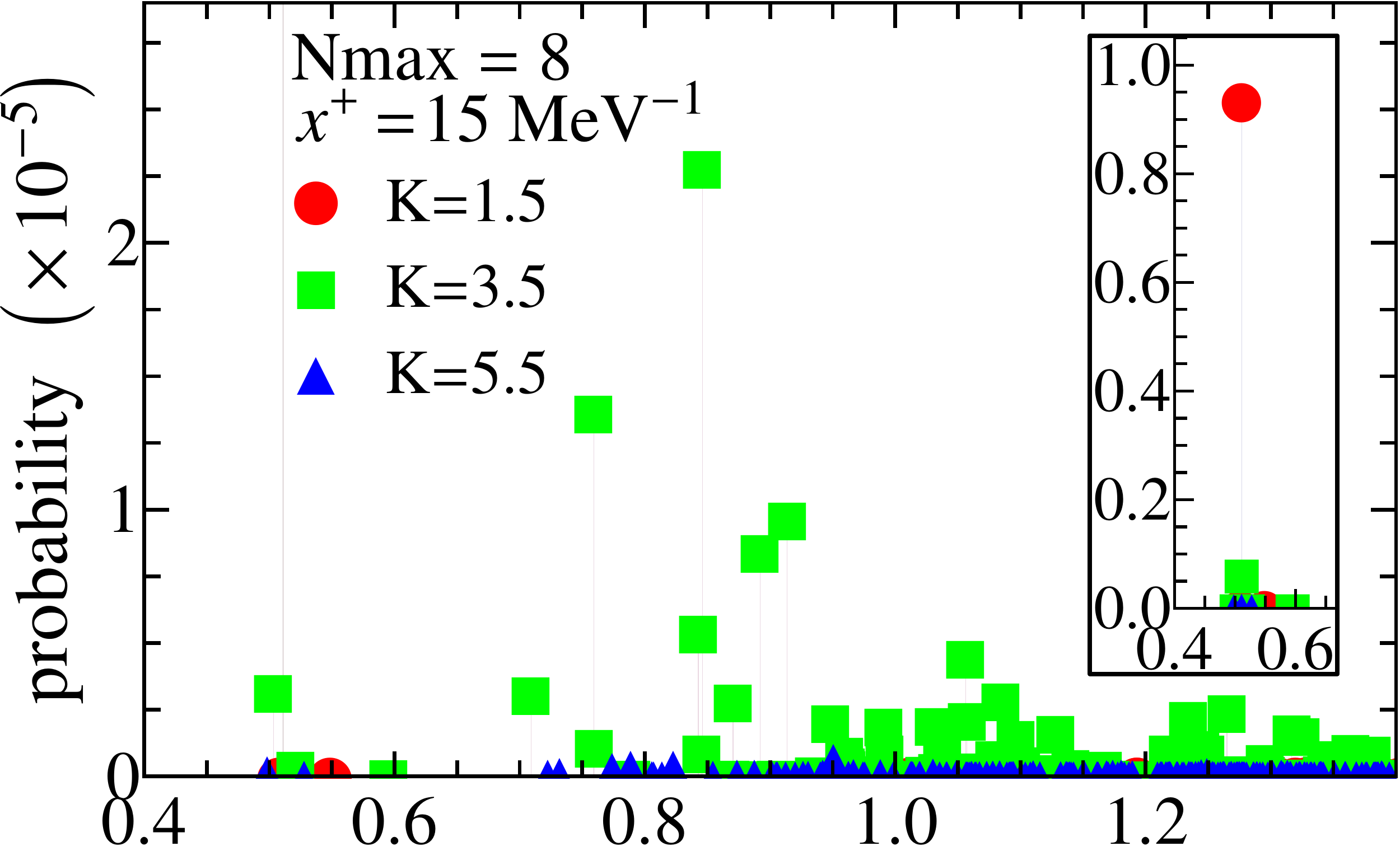} \\
\centering\includegraphics[width=0.45\textwidth]{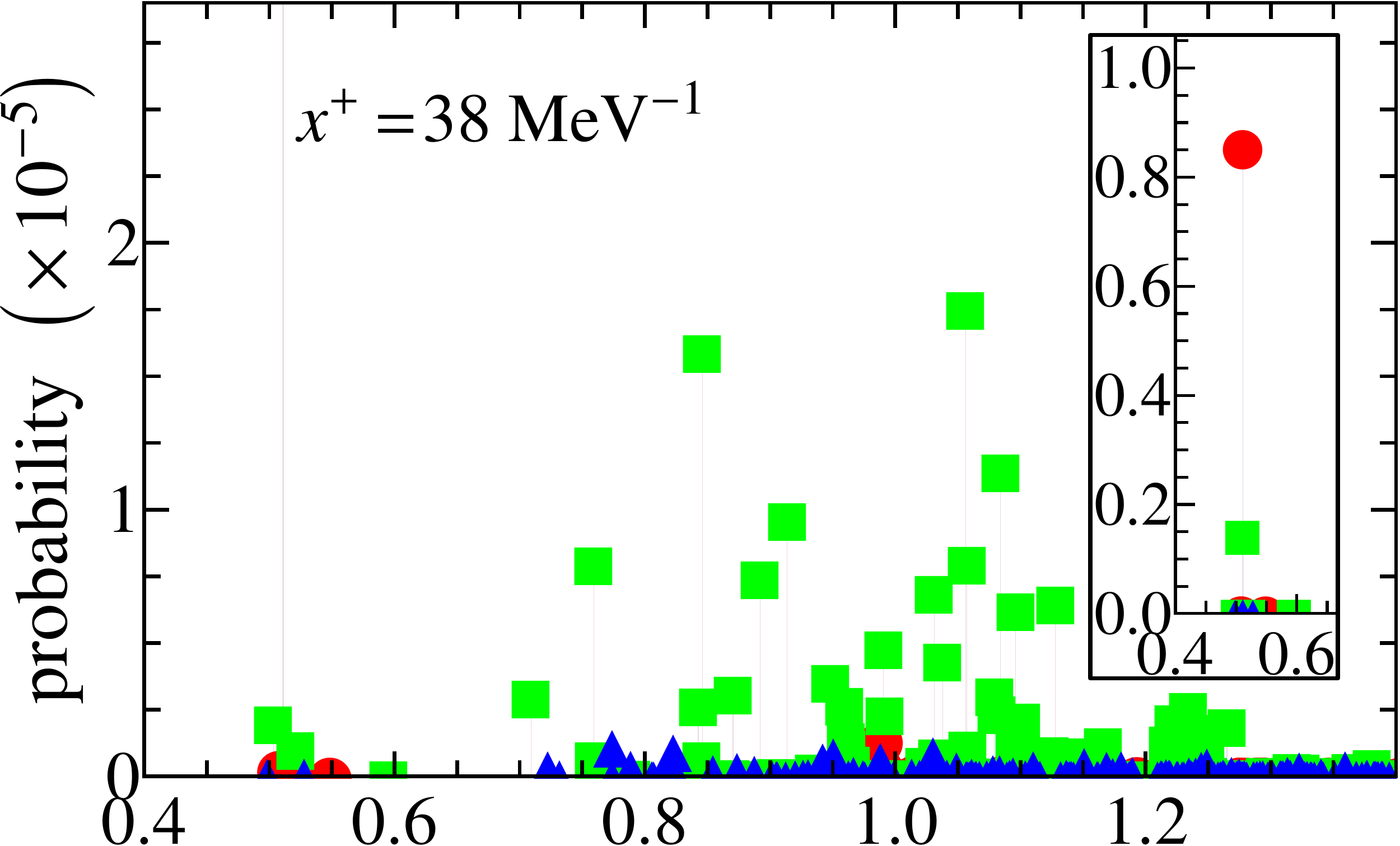} \\
\hspace*{-2mm}\includegraphics[width=0.46\textwidth]{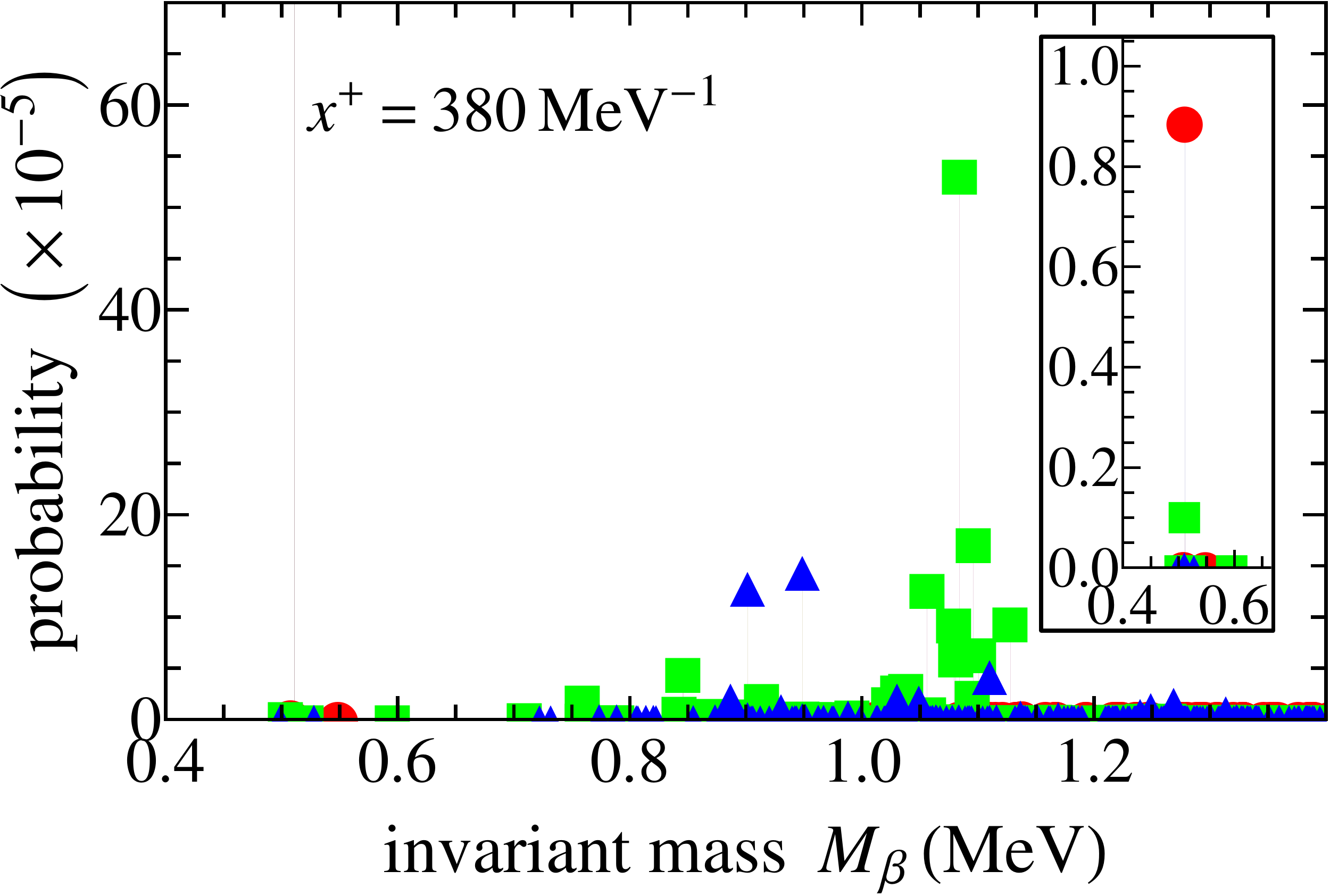}
\caption{\label{fig:state_evol_np} Snapshots of the time evolved system, at $x^+=15,38,380$MeV$^{-1}$ (top to bottom). The background field switched on (off) at $x^+$=0 ($x^+{=}400$MeV$^{-1}$). Each dot represents an eigenstate $\ket{\beta}$ of QED. Horizontal axis: the invariant mass $M_{\beta}$  of the state $\ket\beta$. Vertical axis: probability of finding the state $\ket{\beta}$ in units of $10^{-5}$. The insets show, at normal scale, the (larger) probabilities of finding the single physical electron states (in the $K=1.5, 3.5, 5.5$ segments), with invariant mass $M_\beta=m_e$.}
\end{figure}

We present the evolution of the electron system in Fig.~\ref{fig:state_evol_np}, at increasing (top to bottom) light-front
time. As time evolves, Fig.~\ref{fig:state_evol_np} shows how the background causes transitions from the ground state in the
$K{=}1.5$ segment to other eigenstates of $P^\LCm_\text{QED}$. Both the single electron states and electron-photon states are
populated; the former represent the acceleration of the electron by the background, while the later represent the process of
radiation. Soon after the background is switched on, at $x^+{=}15$\,MeV$^{-1}$, the single electron state in $K{=}3.5$ has been
populated while the probability for finding the initial state begins to drop. Populated electron-photon states appear and form an
initial peak structure, located near an invariant mass of 0.85\,MeV. This initial peak structure consists of the electron-photon states with the largest coupling to the ground state in the $K{=}1.5$ segment.
\begin{figure}[!t]
\includegraphics[width=0.47\textwidth]{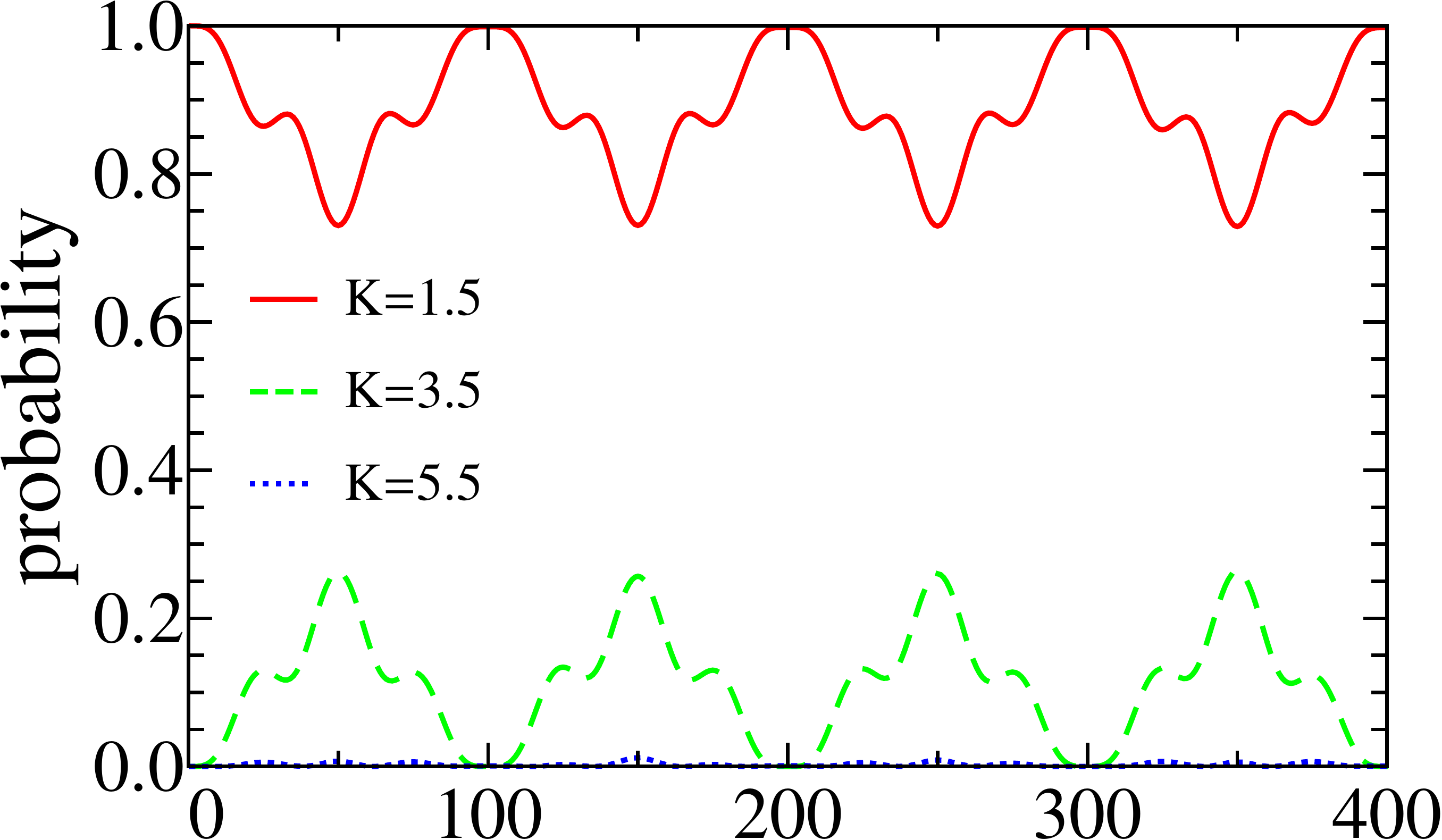}
\hspace*{-2mm}\includegraphics[width=0.48\textwidth]{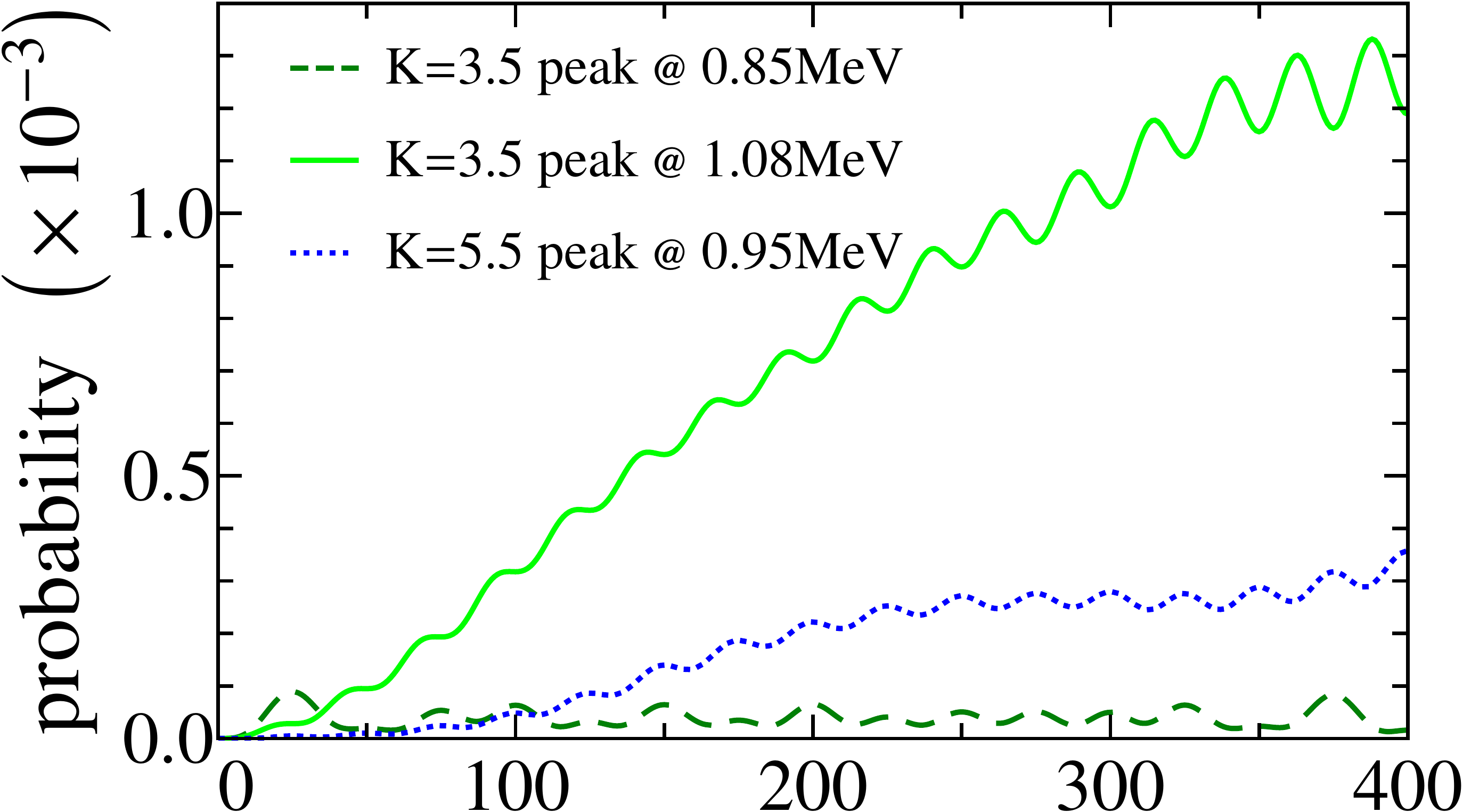}
\includegraphics[width=0.47\textwidth]{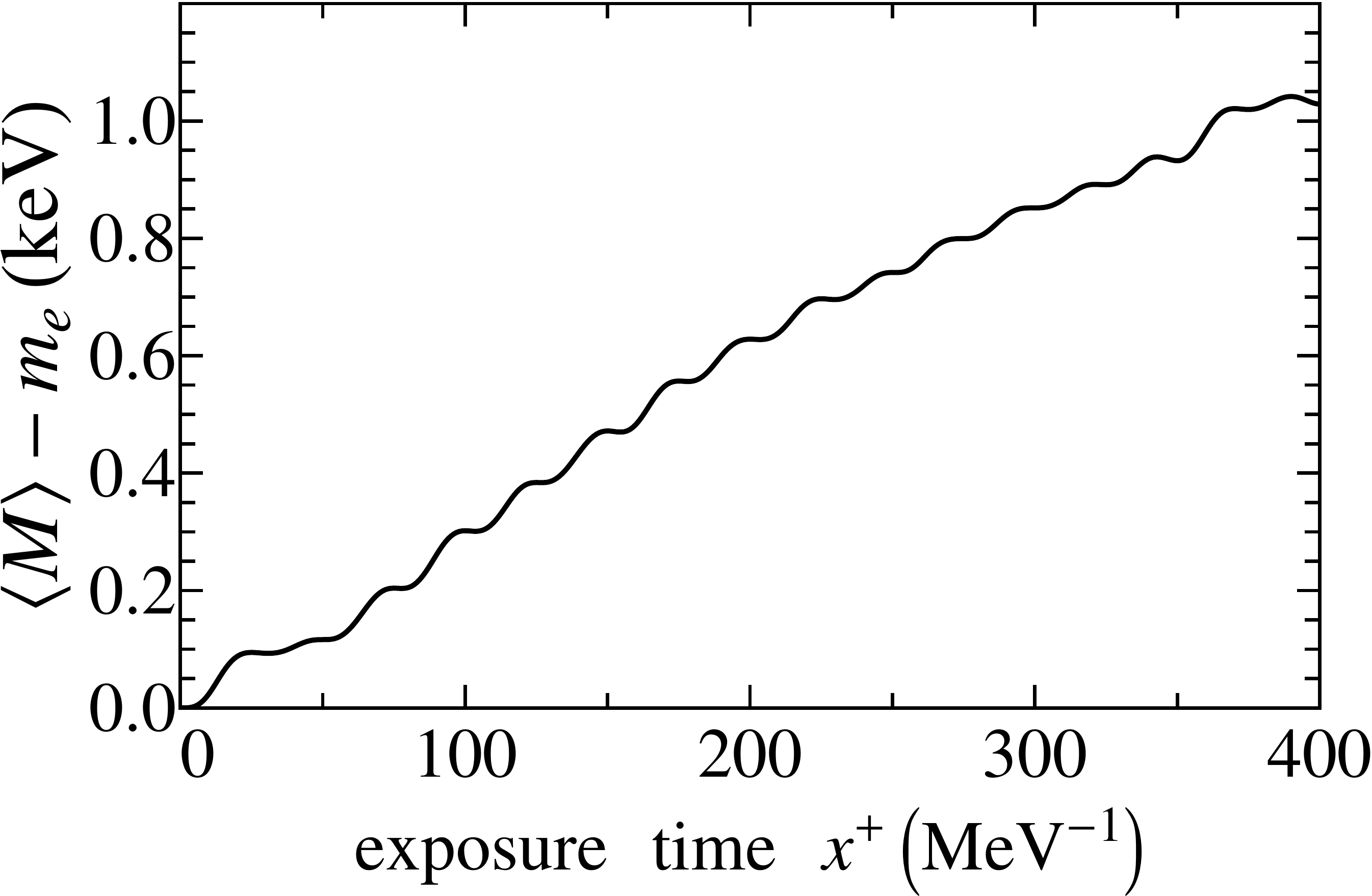}
\caption{\label{fig:prob_v_time} The {\it top panel} shows the time evolution of the single-electron-state probabilities. The periodicity is approximate; radiation losses lead to the {\it middle panel,} which shows the time-dependent probability of produced electron-photon states. Three invariant masses were chosen; the probability shown is the summed probability of all (``binned'') states at the specified $K$-value within a half width of 0.05~MeV around the quoted invariant mass values. The {\it bottom panel} shows the increase of the invariant mass of the system, averaged over all states, as a function of time.}
\end{figure}
According to Fermi's golden rule, most of the integrated transition amplitudes will oscillate as time evolves, leading to oscillating occupation probabilities for the electron-photon states. The exceptions are
those electron-photon states whose light-front energy matches that of the initial single electron plus that transferred to the system by the background. In the $K{=}3.5$ segment, such states form a second peak at an invariant
mass of around 1.1\,MeV, see the middle panel of Fig.~\ref{fig:state_evol_np}. Instead of oscillation, this second peak builds up with time and eventually surpasses the initial peak at $x^+$=0.85\,MeV.

As basis states in the $K{=}3.5$ segment continue to be populated, ``second'' transitions to the $K{=}5.5$ segment eventually become possible.  As a result, a peak in the $K{=}5.5$ segment emerges at the invariant mass of around $0.95$~MeV, see the bottom panel of Fig.~\ref{fig:state_evol_np}. This peak is formed by the second transitions from the $K{=}3.5$ peak states. The fact that the $K{=}5.5$ segment peak is located at a lower invariant mass than that in the $K{=}3.5$ segment (invariant mass 1.1~MeV) demonstrates a case of stimulated de-excitation where, in these second transitions, the electron's $p^+$ is increased, but its momentum relative to the emitted photon ($p_e\!\cdot\!p_\gamma$) is decreased.

In Fig.~\ref{fig:prob_v_time} we present key observables characterizing the main features of the system over the exposure time. The top panel shows the evolution of the probability for finding the three single electron states (in the $K{=}$1.5, 3.5
  and 5.5 segments). As the background field accelerates and decelerates the electron, the transitions between the $K{=}1.5$ and
  $K{=}3.5$ states exhibit an approximate periodic structure. (A very small probability for the $K{=}5.5$ single electron is difficult to see on the chosen scale.) 
  The period for this transition is 100\,MeV$^{-1}$, twice of that for the background, $2\pi/l_+{=}$50\,MeV$^{-1}$. This is
  because the transition frequency is determined not only by the background's $l_+$, but also by the frequency corresponding to
  (the inverse of) the light-front energy spacing between these two states ($\sim$1/100\,MeV in this example). The sum and difference of these frequencies are 3/100 and 1/100\,MeV, corresponding to periods of 33 and
  100\,MeV$^{-1}$, respectively. Therefore, the period of $\sim$100\,MeV$^{-1}$, as the least common multiple of the sum and difference, is seen in the observed transitional between the $K{=}1.5$ and $K{=}3.5$ states.

Another viewpoint of this dynamical situation can be understood as follows: as the electron travels in the background field, it sees periodic structure in {\it both} the $x^+$ and $x^-$ directions, thus the period of its motion is determined by both the frequencies $l_+$ and $l_-$ (the latter via the light-front energy spacing $\Delta P^-{\propto} \frac{1}{K_i}{-}\frac{1}{K_i+k_{las}}$). This periodicity is only approximate, though, because as the electron is accelerated, it radiates photons and thus loses energy (classically, this is radiation reaction).

In the middle panel of Fig.~\ref{fig:prob_v_time} we show the time evolution for the probability of the three aforementioned peaks
in the electron-photon states. The $K{=}3.5$ peak at the invariant mass of 0.85~MeV oscillates with time and illustrates fluctuations in light-front energy consistent with the uncertainty principle. In contrast, the other two peaks accumulate with time signifying
that light-front energy conservation is (approximately) obeyed in the corresponding transitions.

Since the state $\ket{\psi;x^\LCp}$ encodes all the information of the system, it can be employed to construct other
observables. Fig.~\ref{fig:prob_v_time} bottom panel, for example, shows the approximately linear increase of the invariant mass of the system, averaged over all states, reflecting the fact that photons are created as energy is pumped into the system by the background.

\paragraph{Conclusions} We now summarize our results. We have constructed a nonperturbative framework for time-dependent problems in quantum field theory, called ``time-dependent BLFQ" (tBLFQ). Given the light-front Hamiltonian of the system, and an initial state as input, we solve the light-front time evolution of the state. Basis truncation and time-step discretization are the only approximations in this fully nonperturbative approach.

tBLFQ is applicable both to external field problems and to systems  where time-dependence arises from using non-stationary initial states even without external fields.  As an initial demonstration we have applied our framework to ``strong-field QED", and photon emission from an electron excited by a background laser field.

We imagine that another application of this framework would be the propagation of jets produced in relativistic heavy-ion collisions.  For example, from an initial state of back-to-back quark-antiquark jets, the system evolves in the medium of the colliding nuclei modeled as a dissipative background medium with time dependence specified through a model.  The final state population of the leading hadron mass eigenstates could then be predicted.  This application would also involve the non-trivial solution of the hadron spectroscopy as eigenstates in the BLFQ basis.  We are encouraged to speculate in this direction due to the success of the AdS/QCD approach to hadron spectroscopy~\cite{deTeramond:2008ht}, where the basis used is similar to our own.

We acknowledge valuable discussions with K. Tuchin, H. Honkanen, S. J. Brodsky, P. Hoyer, P. Wiecki and Y. Li. This work was supported in part by the Department of Energy under Grant Nos. DE-FG02-87ER40371 and DESC0008485 (SciDAC-3/NUCLEI) and by the National Science Foundation under Grant No. PHY-0904782. A.~I.\ is supported by the Swedish Research Council, contract 2011-4221.

\end{document}